# A Synergistic Approach for Internet of Things (IoT) and Cloud Integration: Current Research and Future Direction

MANZOOR ANSARI, Jamia Millia Islamia
SYED ARSHAD ALI, Jamia Millia Islamia
MANSAF ALAM, Jamia Millia Islamia

Cloud computing and Internet of Things (IoT) have independently changed the course of technological development. The use of a synergistic approach that amalgamates the benefits of both these path breaking technologies into a single package is expected to have flourishing benefits. However, such an integration is faced with numerous limitations and challenges. This paper surveys the different aspects of each of these technologies and explores the possibilities, benefits, limitations and challenges that rise from the development of a convergent approach. We have also investigated the current research and future direction.

## 1 INTRODUCTION

The Internet of things and Cloud Computing are two innovative paradigms that are an integral part of our daily lives and are drawing in huge enthusiasm from both industry and the academic world. Cloud of Things (CoT) is an enthusiastic view of the IoT paradigm in which every useful device, namely ' intelligent things' are fully Internet- connected and cloud-based [75]. The IoT is expected to grow to 35 billion units by 2020, making it one of the primary "Big Data" source s with features such as volume, velocity, value, density and heterogeneity y [64].

The primary objectives of IoT are to provide network infrastructure with inter operable communication protocol and software to facilitate connection and incorporation. IoT is an Internet -based physical system embedded with operating systems, devices, actuators, sensors ,electronic components, and wireless connectivity to collect and exchange information on these objects [52].

Salient features of an IoT system are as follows:

• IoT portrays a system in which objects in the physical environment or devices inside or connected to these objects are associated to the Internet by wireless or wired Internet connectivity.

• Various kind s of local area network-access technologies like RFID, Zigbee, Bluetooth, NFC and Wi-Fi are used by these IoT sensor devices.

• Sensors can also be connected to a wide area such as long-Term Evolution (LTE), Global System for Mobile communication (GSM), General Packet Radio Service (GPRS) and 3G services.

Authors' addresses: Manzoor Ansari, Jamia Millia Islamia, New Delhi, manzoor188469@st.jmi.ac.in; Syed Arshad Ali, Jamia Millia Islamia, New Delhi, arshad.ali71@gmail.com; Mansaf Alam, Jamia Millia Islamia, New Delhi, malam2@jmi.ac.in.

• IoT system can be divided into three components that support the efficient functioning of the system.

• Presentation
• Middleware
• Hardware

Cloud computing, on the other side, has almost infinite storage and system capabilities that can resolve most of the IoT issues. Thus, a novel IT paradigm that incorporates both the Cloud and IoT technologies is known as Cloud of Things (CoT), that enhance the present and future of IoT system and manage IoT devices and their resources in an efficient way. The Cloud of Things is an emerging technological paradigm that is used by most of the industries and manufacturers to enhance their productivity and make the system more efficient [29]. The researchers have discussed their work [62] that Cloud computing can be use as an apt platform for analytics of big data in case of the data storage and computing needs. In the study of a research [34], it is identified many challenges and needed improvement in coming days for efficient energy.

## 1.1  Internet of Things

In 1999, British technology pioneer Kevin Ashton coined the word "Internet of Things (IoT)" to describe a network system where sensors can connect objects with the internet in the physical environment. In IoT, "things " are related to any object on the face of the earth, whether it is a communication or a non-communication device.

Although IoT is universally accepted, no global definition of the term exists. The following are different IoT definitions:

***"The term Internet of Things belongs to scenarios in which connectivity of the network and computing capability strikes to objects, sensors, and usual items, not personal computers, granting these devices to create, interchange, and utilize data with minimum human involvement".***

According to the Oxford Dictionary [13], *" The interconnection of electronic systems embedded with everyday objects over the Internet, enabling them to send and receive information ".*

The RFID group describes as: *"The worldwide network of interconnected objects uniquely addressable based on standard communication protocols".*

### 1.1.1  *Layered Architecture of IoT.* IoT layered architecture is given in Fig.1. There are five layers namely Perception, Network, Middleware, Application, and Business layers [73]. The last three layers of this architecture exist on the cloud [52].

• **Perception Layer:** The perception layer is also known as the "Sensing Layer" or "Physical Layer" in IoT. The purpose of this layer is to acquire the data from the physical environment with the help of sensors and actuators. This layer identifies, collects, processes and then transmits information to the network layer. In the physical environment, it senses certain physical parameters or identifies other smart objects. It also collaborates with the IoT node on local and short-range networks.

• **Network layer:** The IoT network layer provides the function of routing and transmitting data over the Internet to various IoT hubs and devices. Cloud computing platforms, Internet gateways, switching and routing devices etc. operate on this layer using some of the latest technologies like Wi-Fi, LTE, Bluetooth, 3G, Zigbee etc. The network gateways act as the mediator between different IoT nodes by aggregating, filtering and transmitting data from and to various sensors. It is the responsibility of the network layer to connect to other intelligent things, network devices, and servers. Its characteristics are also used to transmit and process sensor data.

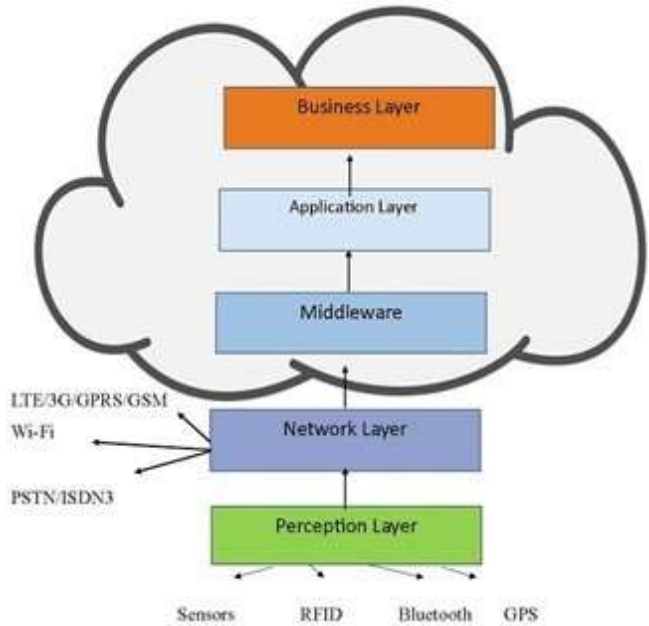

Fig. 1. **Layered Architecture of IoT**

- **Middleware:** The middleware layer is also known as the "processing layer". It analyses, processes and stores, enormous amounts of data from the Network layer. It can perform and provide the lower layers with a variety of services. It uses numerous technologies, such as databases, cloud computing, and modules for big data processing.
- **Application Layer:** The application layer is responsible for providing the user with specific application services. It describes different applications, such as environmental monitoring, smart homes, smart farming, smart cities and smart health care etc. Data authenticity, integrity and confidentiality are ensured by the application layer. The aim of this Layer is to achieve a smart environment.
- **Business Layer:** The business layer intends the entire IoT network, along with applications, business and revenue models and privacy for users.

## 1.2   Cloud Computing

There are several definitions proposed for cloud computing. According to NIST: *"Cloud computing offers a model for convenient on-demand network access to a shared pool of configurable computing resources that can be accessed and delivered on time with minimal management or service effort"* [65]. Cloud computing can be treated as a long-held delusion about figuring as a utility and can possibly change an enormous piece of IT business, forming the way IT equipment is planned and acquired [37]. In some research work big Data scientific work flow in cloud execution and deployment is an area that needs research thinking before a synergistic model [61].

The following functionality has been suggested by cloud computing such as on-demand self-service, extensive network access, resource pooling, rapid elasticity, and measured resources [51] [3] [9]. Cloud characteristics have been summarized in Fig. 9.

- **On-demand self-service:** Several Cloud computing services can be distributed from the service provider without human interaction. Such services are storage capacity, instances of virtual machines and instances of databases, in addition to several others. Production enterprises can use a self-service web portal as a platform for accessing their cloud accounts to see their cloud services, their use, as well as delivering or de-providing services as required. Consumers can provide computing capabilities as required, such as server time and network storage, without human interaction with each service provider.
- **Extensive network access:** Cloud services are available on the network, and various user applications may access them. It prefers strong broadband communication networks including Internet or a local area network (LAN) in the case of private clouds. The bandwidth and latency of the network are important features of cloud computing, as they contribute to the quality of network service (QoS). There are numerous network-accessible functions that can be accessed through conventional processes to support the use of thin or dense heterogeneous customer systems (e.g. portable devices, tablets, laptops, and workstations).
- **Multi-tenancy and Resource pooling** : A multi-tenant system is supported by cloud computing resources. Multi-tenancy supports multiple clients who maintain their data security and privacy to share the same software or physical infrastructure. System resources such as processing, memory, space and bandwidth of the network are distributed to handle different users that used a multi-tenant system where various types of virtual resources are dynamically allocated and relocated as required. Resource pooling means, several customers have the same physical resources. The resource pool of providers should be highly big and versatile enough to satisfy various customer demands and ensure economical scale. Resource allocation must not impact the performance of critical manufacturing applications when it comes to resource pooling. Scheduling of resources are important aspect in cloud computing, a research [35] explained present scenario of algorithm for Task Scheduling algorithms based on parameters of various scheduling.
- **Rapid elasticity and scalability:** Cloud computing resources may rapidly increase or decrease and, in some cases, automatically respond to business requirements. It's a vital characteristic of cloud computing. Availability, efficiency and therefore costs can be adjusted up or down without any additional contract or fine. Elasticity is a milestone in cloud computing and ensures that any cloud-computing resource can be produced and delivered quickly by production organizations. In case of storage or virtual machines or customer applications, rapid provisioning and de-provisioning can apply. Scalability is more pragmatic and liberal.
- **Measured service:** The applications of cloud computing services are measured, and production companies pay for what they have used accordingly. The authors of a research [74] have discussed the problems which is faced by researchers work in the field of bioinformatics in order to intend their research in minimum expenditure and quick way can be solved in easy way using Cloud computing .Cloud technologies automatically govern and optimize resource use by leveraging metering capabilities at certain abstraction levels, such as processing, storage, bandwidth and active user accounts. The need for resources for both the provider and the consumer can be supervised, measured and conveyed in a transparent manner. The characteristics of cloud computing is shown in fig.2.

## 2   GOOGLE RESEARCH TREND FOR IOT AND CLOUD COMPUTING

In one of the research it is shown that production kind trace similar to the ones in cloud environment [32]. In the last few years, we have reviewed the affluent and eloquent literature in this field. In Fig. 3, the popularity of both technologies according to the google research trends has been shown. The growing research trend of google from the year 2014 to present 2019, according to the comparison of IoT and Cloud Computing is given in Fig. 4.

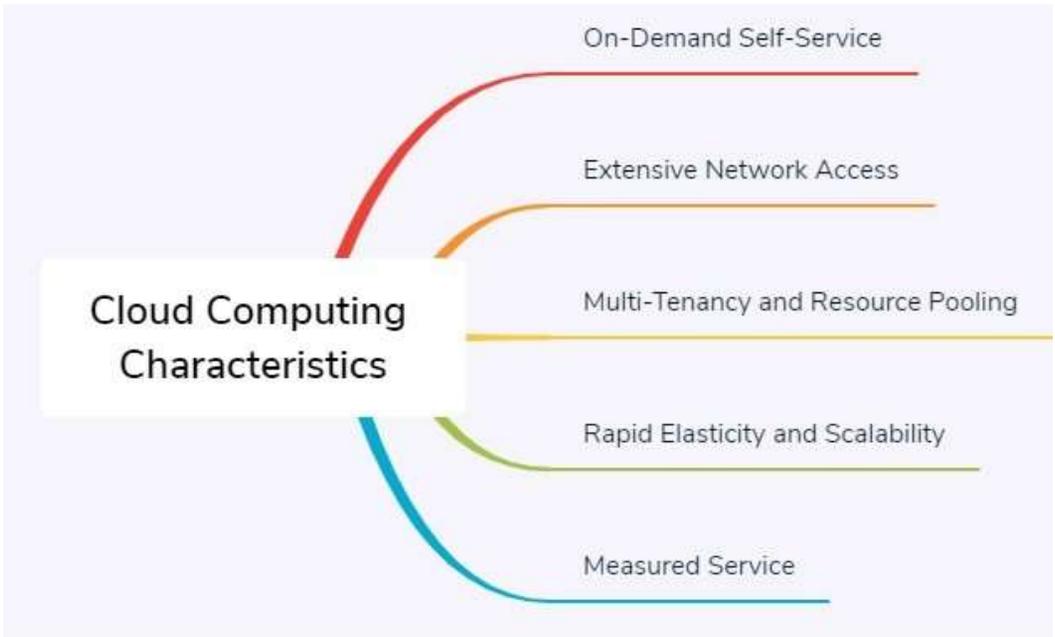

Fig. 2. **Cloud Computing Characteristics**

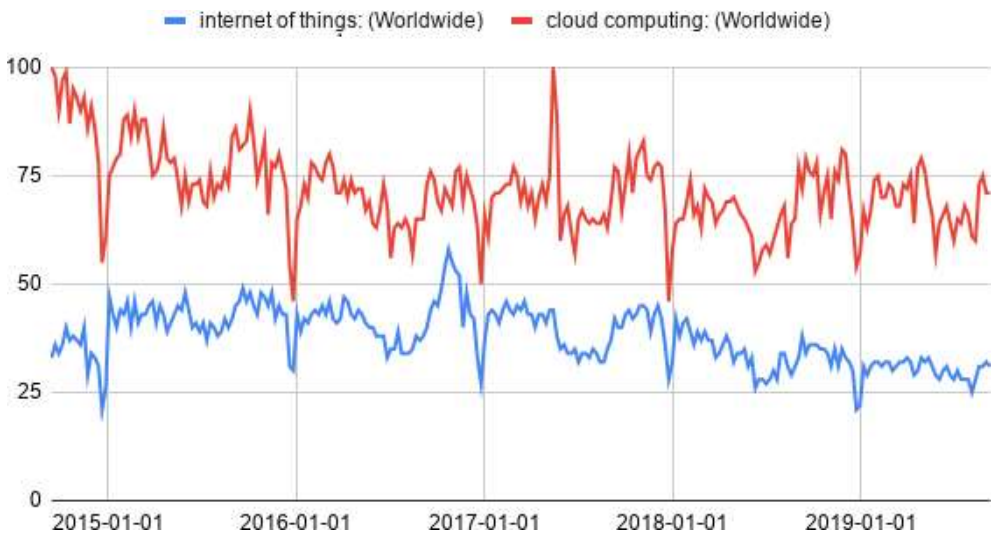

Fig. 3. **According to Google Research Trends (Comparison IoT and Cloud).**

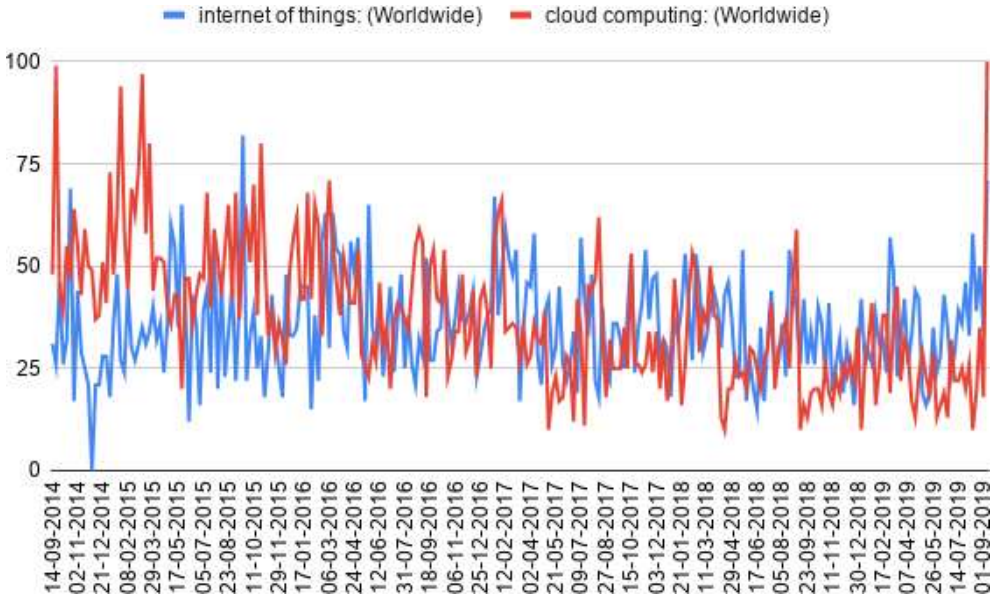

Fig. 4. **According to Google research trends (comparison IoT and cloud-based on Academic Conferences & Publications worldwide)**

## 3    ANALYSIS OF CLOUD AND IOT RESEARCH ON ELECTRONIC DATABASES

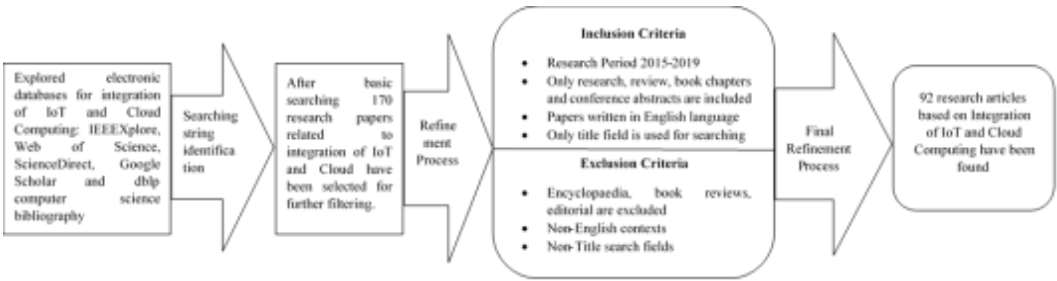

Fig. 5. **Research Methodology**

Many research papers have been written for IoT and Cloud Computing separately. These research articles highlighted the IoT and cloud computing issues and challenges. Some researchers have proposed solutions for these challenges and issues.

For the investigation of the integration of IoT and Cloud Computing technology and in order to answer the following questions, various electronic databases have been explored for searching the research articles related to IoT and Cloud Computing integration. These research questions are:

**RQ 1.** How Cloud is associated with IoT technology?

**RQ 2.** What are the benefits of Cloud-based IoT?

**RQ 3.** What are the various complementary approaches of Cloud -based IoT?

**RQ4.** What is the various Cloud-based IoT platforms, middleware technologies and architectures?

**RQ 5**. What are the applications of Cloud-based IoT technology?

This research explores on Web of Science, ScienceDirect, IEEEXplore, Google Scholar and dblp computer science bibliography electronic databases for searching research articles related to IoT and Cloud Computing integration.

Fig. 5 shows the steps involved in the selection process of the relevant research articles (inclusion and exclusion criteria). In the first step, various electronic databases ( IEEEXplore, Web of Science, ScienceDirect, Google Scholar and dblp computer science bibliography) were explored for searching the articles related to IoT and Cloud Computing integration.

The database s were searched with search string such as (("IoT" OR "Internet of Things") AND (" Cloud Computing ")) for the title field. The combined databases returned 170 articles without any refinement applied to the search data.

Then the refinement process was done for inclusion and exclusion of following conditions, which are mentioned in Fig. 5. After refinement process, 92 research articles were selected that consisted of papers from the year 2015 to 2019 and all research articles, review, book chapters and conference proceedings written in the English language were selected for the final review process.

These 92 papers have research related to IoT and Cloud Computing; in which some papers are related to IoT and Cloud Computing integration as well, while others are based on Cloud of Things platforms, benefits and challenges of CoT, CoT applications, architectures and middleware technologies.

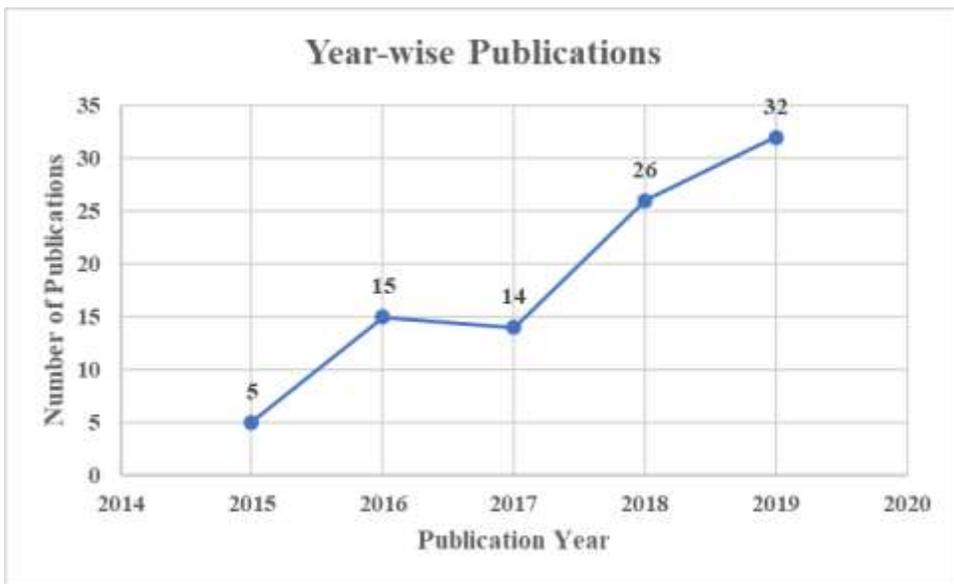

Fig. 6. **Year-wise Publication of CoT related research papers**

The searched dataset consists of 92 research papers from different databases, which have been analyzed and shown in Fig. 6 according to the year of their publications.

## 3.1    Distribution of Research Documents

After exclusion of encyclopedia, book reviews and editorial materials, only 92 research articles have remained in which 81% papers are research articles, 13% are mini-reviews, 4% are major review articles and only 2% are meeting abstracts. Fig. 7 shows the percentage distribution of research document-type. The dataset also describes the quality of the research papers. The papers published in reputed journals are considered good research papers.

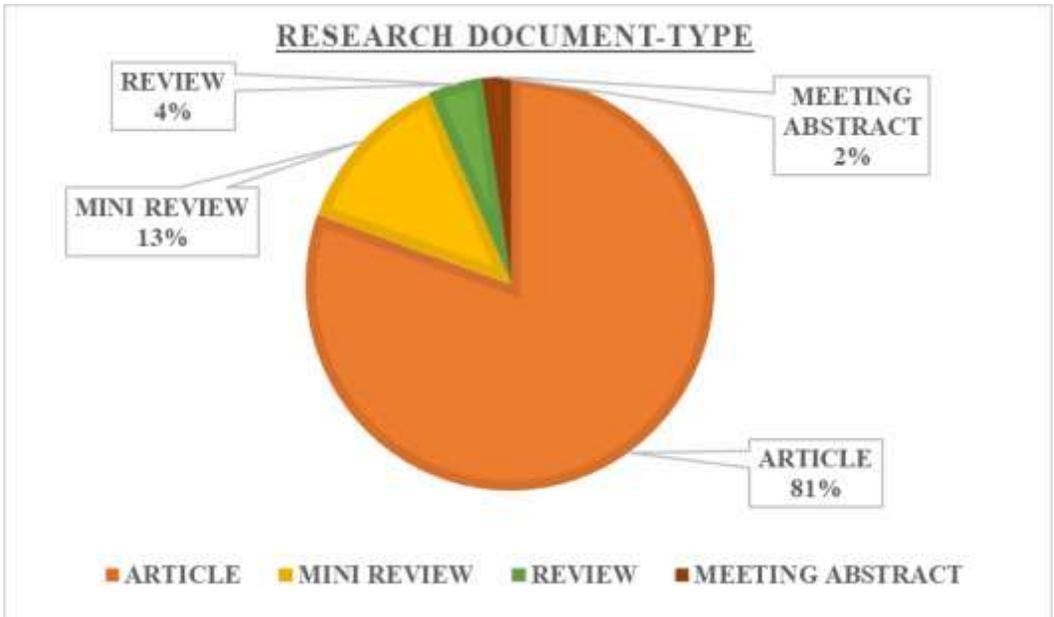

Fig. 7. **Distribution of Research Document-type**

Fig. 8 shows the distribution of papers according to the reputed publishers in which these papers have been published. Among the published papers, IEEE consitutues 38% of papers, 17% papers are from Springer, 9% papers are from ACM and Elsevier both and rest of the 27% papers are from various other journals.

Fig. 9 describes the year-wise research article publications from the year 2015 to the year 2019 in the reputed journals. These papers are further studied and used to discuss the upcoming sections of the paper. The remaining of the paper is organized as follows. Section 1 discusses the Internet of things, IoT architecture layer and illustrates cloud computing and its characteristics. Section 2 explained google research trends for IoT and Cloud, Section 3 focuses on analysis of cloud and IoT research on electronic databases and distribution of research documents, section 4 overview of related studies, section 5 discusses fusion IoT and cloud, section 6 focuses on future of IoT and its interaction with blockchain, CoT challenges and open research issues, section 7 enlighten conclusion and future direction.

## 4   OVERVIEW OF RELATED STUDY

Recently, several studies on IoT and cloud-based technologies have been explored in a broad spectrum of technical aspects. Lots of efforts have been made to offer review articles on this area of research in various fields. These articles [42] [45] [44] presented a review of recent efforts to adopt Cloud Computing in various IoT scenarios and applications.

Botta et al. [42] , reviews literature to define complementary approach of cloud and IoT as well as the primary drivers for their integration into a unique environment. As the acceptance of the CloudIoT paradigm allowed several unique applications, for each of them, the main research challenges are extracted. To define current directions for research, they further analysed such challenges. Lastly, comparing their main aspects, affordable platforms and projects identified open challenges and directions for future research in this field.

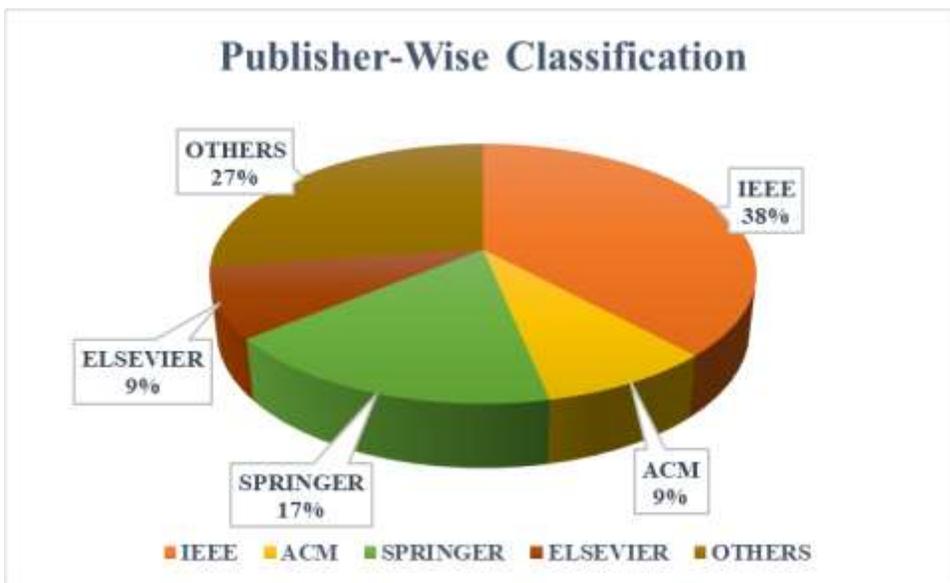

Fig. 8. **Distribution of Papers according to Publishers**

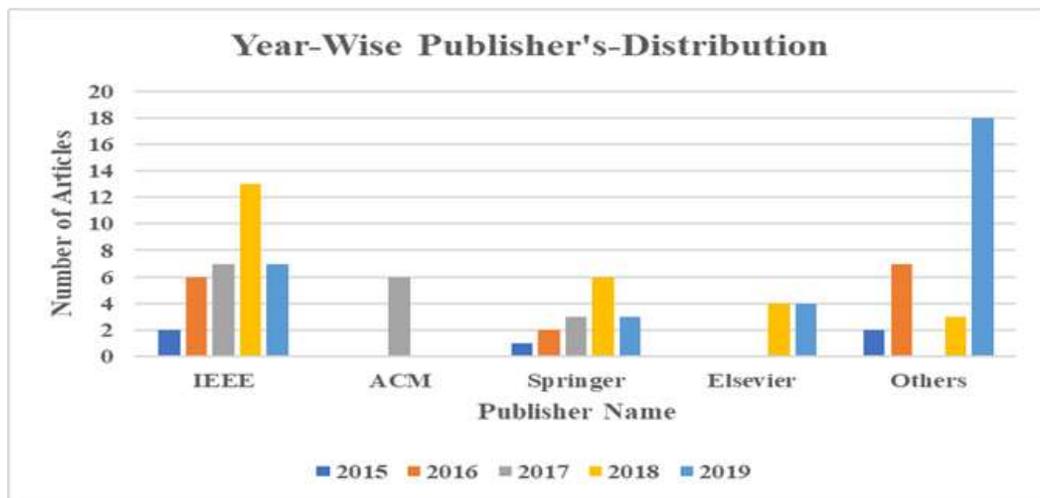

Fig. 9. **Year-wise distribution of Papers according to Publishers**

Díaz et al. [45], presents an integration component study of cloud platforms, cloud infrastructure and middleware IoT technology. Also, numerous data analytical techniques and approaches for integration are discussed along with various research challenges and issues.

Dang et al. [44], evaluate new IoT products, health sector implementations, patterns, as well as assess the latest IoT and cloud computing frameworks published since 2015. Researchers discuss emerging innovations such as cloud, living assistance, big data, and wearable technologies and explore how they help to establish Internet of things (IoT) and cloud technologies sustainably in the medical sector, through the implementation of various Internet of things (IoT) and e-health regulations and policies across the world. In addition, several (Internet of things) IoT security and privacy issues are thoroughly reviewed, consisting of risks, security and threat measures. Ultimately, this paper explores well-known models for assessment of security risks to tackle and present trends, opportunities, and challenges for future development in IoT-based healthcare.

Gill et al. [50], investigates how three developing paradigms (IoT, AI and Blockchain) affect future cloud computing techniques. They also discuss various technologies that help these concepts and encourage researchers to address the existing status and research direction of cloud computing. Finally, a computational cloud futurology framework was developed to explore the effect on cloud system development of evolving paradigms. Table 1 summarize the year of publications, the main topic, and contributions from previous related surveys on the integration of IoT and cloud technologies, as well as key contributions for our review document.

## 5  FUSION OF IOT AND CLOUD

The two universes of IoT and Cloud technologies have seen independent development and maturity cycles. These technologies are enticing and diverse in terms of research domains. The integrated paradigm does not have no standard name. Some generic names have been identified in literature such as CoT, Cloud of things, Web of Things (WoT), Cloud of Everything, Internet of things Cloud (IoT Cloud) [36]. Fig 10 shows the fusion of IoT and cloud technologies.

On one hand, IoT can leverage the virtually unlimited capacity and resources of Cloud to compensate for technical parameters (such as processing, storage, power). In particular, the cloud computing can provide an effective solution to implement the management and composition of IoT services, as well as applications that leverage the things or data they produce. On the other hand, the cloud can leverage IoT by expanding its range to address real-world issues in a more dispersed and vibrant way and providing new installations in a wide variety of real-world situations [42].

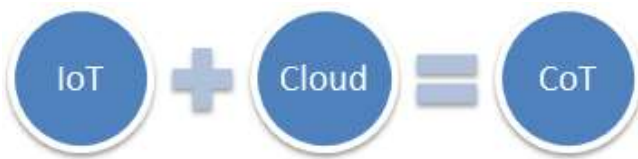

Fig. 10. **Fusion of IoT and Cloud**

IoT consist of devices with low processing power and storage because of their limited vigorous abilities, a contrast to the various complex processing tasks that need to be carried out. To reduce this limitation, such devices could act as simple data providers and send data directly to the cloud for processing and storage [43].

Cloud computing acts as an intermediate layer between things and applications, hiding all the complexity and functionality needed to implement it. This paradigm will affect the future development of applications where data collection, analysis and distribution would produce new challenges to be tackled in a multi-cloud environment [42]. The research methodology adopted in this work is shown in Fig. 11.

### 5.1  Complementary approach IoT and Cloud

Cloud computing and IoT work to improve the efficiency of everyday tasks and both have a complementary relationship. IoT generates a lot of data on the one hand, while cloud computing paves way for this data to travel on the other hand. The reciprocal parameter of IoT and Cloud that emerge from the different literature proposals and inspire the CloudIoT paradigm are revealed in Table 2.

Table 1. **Comparison of existing survey papers on IoT and related technologies**

| Research Articles | Year of Publication | Topic | Main Contributions |
|---|---|---|---|
| [42] | 2016 | **"Integration of Cloud Computing and Internet of Things: A survey"** | A review on IoT and Cloud technology, complementary approach, applications, challenges, platforms, current projects and open issues. |
| [45] | 2016 | **"State-of-the-art, challenges, and open issues in the integration of Internet of things and cloud computing"** | A comprehensive survey of underlying platforms, infrastructures, middleware, integration proposals, data analytics technique, challenges, open issues and future directions. |
| [44] | 2019 | **"A Survey on Internet of Things and Cloud Computing for Healthcare"** | A brief review of IoT and Cloud Computing in Health-care that analyse IoT component, applications, market trends. Also discuss various threats, vulnerabilities, attacks, security models, government policies and challenges in healthcare. |
| [50] | 2019 | **"Transformative Effects of IoT, Blockchain and Artiftcial Intelligence on Cloud Computing: Evolution, Vision, Trends and Open Challenges"** | A concise overview to discuss the impact on the development of cloud computing of three developing paradigms (IoT, A.I and Blockchain). |
| This paper | 2020 | **"A Synergistic Approach for IoT- Cloud Integartions"** | A systematic review on the Integration of Internet of things (IoT) with Cloud Computing with detailed discussion on concepts, complementary approach architecture, applications, middleware technologies, Existing CoT Platforms, Cloud-based IoT applications, Advantages and futurology Internet of things (IoT) with Blockchain. |

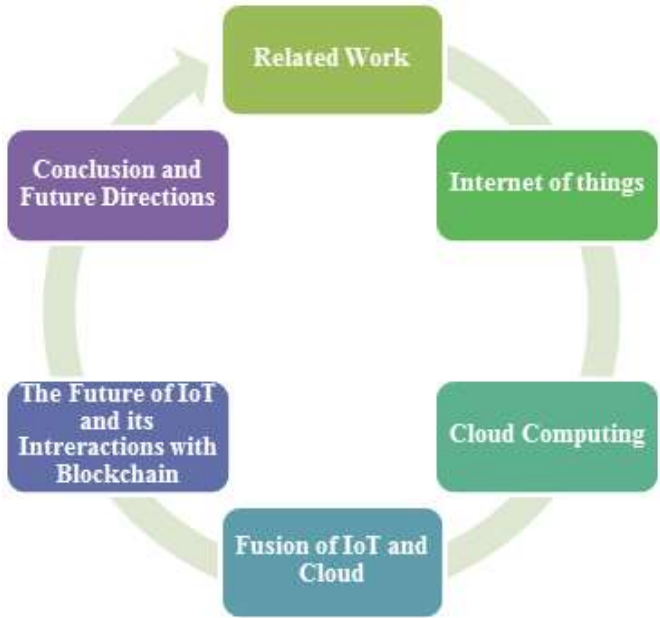

Fig. 11. **Research methodology adopted [42]**

Table 2. **IoT Vs CLOUD COMPUTING**

| Parameters | IoT | Cloud |
|---|---|---|
| *Characteristics* | Pervasive | Centralized |
| *Storage capabilities* | Limited or none | Unlimited |
| *Big Data* | Source | Means to manage |
| *Connectivity* | Internet as a point of convergence | Internet for service delivery |
| *Computational capabilities* | Limited | Virtually unlimited |
| *Reachability* | Limited | Ubiquitous |
| *Components* | Real world objects | Virtual resources |

## 5.2  IoT Middleware Technologies

Between Application Layer and Perception Layer, there is a Software Layer that is called Middleware. Frequently encountered problems like dependability, heterogeneity, inter-operability and security can be handled by this Software Layer. According to Farahzadi et. al [49] Middleware can be considered as a "network-oriented" vision. The primary objective of the middleware technology is to provide the abstraction and interaction functionality of the devices in IoT implementation in order to achieve an inescapable integration with other technologies such as cloud services [73]. IoT-based Middleware overall concept and main characteristics or features are shown in Fig. 12.

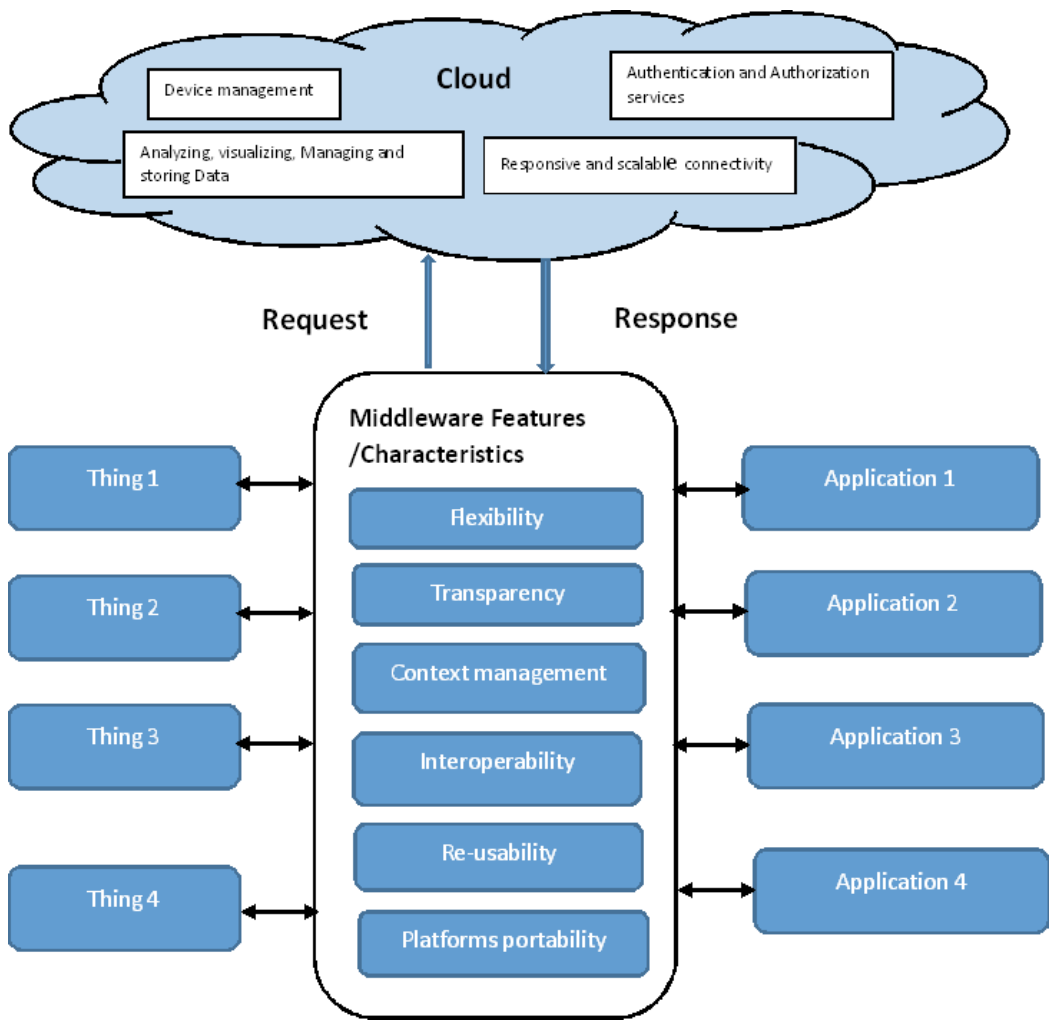

Fig. 12. **CoT- based Middleware -overall concept and main characteristics /features [49]**

### 5.3   Existing platforms for Integration

IoT cloud platforms bring together capabilities of IoT devices and Cloud Computing delivered as a service over an end-to-end platform [58]. Several IoT cloud providers are currently emerging into the market to leverage suitable and specific IoT-based services [70]. Each of the determined platforms in the short-listing process is a highly accepted solution in the market [54]. This paper discusses the current market to select the top -listed IoT platforms. This analysis can help in providing the most optimal environment for running IoT applications. This section provides detailed knowledge about the existing IoT cloud service providers and their pros and cons in concrete form. The architecture and application of top ten middleware is listed in table 3 and IoT platform is given in table 4.

Table 3. **Architecture and Applications of top ten Middleware**

| S. No. | Middleware | Applications | Architecture | Cloud Based |
|---|---|---|---|---|
| 1. | **Aura** [5] | Pervasive computing environment | Distributed | No |
| 2. | **ABC &S** | Car parking automation | Based on Service | Yes |
| 3. | **Carriots** [21] | Smart city, Smart energy | Based on Service | Yes |
| 4. | **CARISMA** | Mobile computing | Distributed | No |
| 5. | **DropLock** | Smart Home Deployment | Based on Service | Yes |
| 6. | **OpenIoT** [17] | Smart cities and mobile crowd sensing | Based on Service | Yes |
| 7. | **Rimware** | Heart Rate Monitor (HRM) and Smart Lighting (SL) | Based on Service | Yes |
| 8. | **ThingWorx** [24] | Agriculture, Smart cities and smart buildings | Based on Service | Yes |
| 9. | **VIRTUS** [26] | E-Health | Distributed | No |
| 10. | **XIVELY** [? ] | Home appliances connectivity and management | Based on Service | Yes |

Table 4. **IoT Platforms**

| Platform | Features | Pros | Cons |
|---|---|---|---|
| **KAA** [14] | • It is an opensource IoT platform for deploying, managing and monitoring Cloud-based IoT applications. It provides lightweight IoT connection protocol, devices and things are identified by digital twins.<br>• It provides scalability, data analytics and visualization and reliability of the system. | Applications related to NoSQL and Big Data are implemented | It supports less heterogeneous hardware devices |
| **Carriots** [16] | • It is a cloud-based platform that helps enterprises to build their IoT application very easily and in a cost-effective way.<br>• It is based on a Cloud-based platform model that enables users to manage the device remotely, enable alarm and data export features. | It supports the trigger-based application | Complicated user interface |

*Table 4 continued*

| | | | |
|---|---|---|---|
| **Temboo** [22] | • Temboo is an IoT application development platform based on Cloud, which is more focused on the smart inventory management for production industries.<br>• Smart sensors are used to monitor the physical assets, give an alert message and generate reliable and useful information. | Choreos application are implemented easily | Resources intensive applications are not supported. |
| **SeeControl IoT** [20] | • It is a Cloud-based service for the IoT device management and communication among them.<br>• It also provides a push/pull architecture based on open API to analyse ad visualize sensor data. | Messaging service between the devices using Push/Pull messaging protocol | Data visualization not provided efficiently |
| **SensorCloud** [2] | • It is a Cloud-based platform for Smart IoT device's data storing, visualizing, monitoring and analysis.<br>• MathEngine and FastGraph tools are used for data analytics and data visualization respectively. | It can manage a huge amount of resources | It is difficult to connect open source devices |
| **Etherios** [11] | • Etherios is now combined with the Westmonroe partners. It provides a wide range of product and services.<br>• It is based on the Cloud PaaS model to facilitate many enterprises to connect their products and monitoring them in real-time. | For third party devices and special devices, cloud-based services have provided | Selection of devices are restricted |
| **Xively** [? ] | • Xively is now a part of Google Cloud Platform product family.<br>• It provides several application-specific sub-products like Watts, Lutron, Sato, Freight farm, Tekmar and SureFlap. | Integration with devices is very easy | Poor management of the system |
| **Ayla's IoT cloud fabric** [7] | • Ayla is an IoT platform for IoT application development. It has featured like Edge computing, device management and application development.<br>• Any device can be connected very easily and in a cost-effective way.<br>• For end-to-end support in mobile and connected devices, it provides an effective software agent. | A mobile-based application can be easily developed. | Small scale development is not supported |

*Table 4 continued*

| | | | |
|---|---|---|---|
| **Thethings.io** [23] | • An effective API is provided for back-end support for IoT developers.<br>• Due to hardware uncertainty, it can connect only those devices that support HTTP, MQTT and CoAP protocols. | Device uncertainty support | Third-party development is not easy |
| **Exosite** [12] | • It is an IoT software development platform that can connect enterprise devices to the market.<br>• Real-time data analytics and visualization provided to the user using Cloud-based Software as a Service (SaaS). | Easy installation of the system | Lack of Big data provisioning |
| **Arrayent Connect TM** [1] | • Heterogeneous IoT devices and systems can easily be connected to the enterprise's smart devices and web applications. | Supports elasticity. | Lagging in trigger-based services due to system configuration |
| **Open remote** [18] | • The user can join any device, protocol and system design using any available resources.<br>• A user can develop any customized software application using remote Cloud's services. | It supports open Cloud services | Insufficient data management |
| **Arkessa** [4] | • It provides connectivity and management of devices for enterprises. | Enterprise based design model | Not enough visualization tools |
| **Axeda** [6] | • IoT and M2M applications are implemented using Cloud-based designing products and tools. | Management of data is based on the M2M model | Third-party service dependency |
| **Oracle IoT cloud** [19] | • It provides four important features: open, insight, secure and acceleration.<br>• It can connect any kind of devices with end to end security to provide business values with minimum risk. | It supports data managements | A problem in the connectivity of open source systems because of size limit |
| **Nimbits** [15] | • It is a hybrid Cloud model that provides IoT services using Edge Computing in constrained enables the system.<br>• It also filters the Cloud data and uploads to the Cloud server efficiently. | Develops can use it easily without any problem | Real-time data processing is poor |

| *Table 4 continued* | | | |
|---|---|---|---|
| **ThingWorx** [25] | • It provides a data-driven decision-making platform.<br>• Search-based intelligence provided by the SQUEAL model for searching, querying and analysing data in M2M and IoT based applications. | Application-based on data intensity can easily develop. | Restriction in many devices simultaneously |

## 5.4 Benefits of Integration IoT and Cloud:

IoT environment consists of sensors, actuators, network devices and storage devices. Due to limited processing and storage capacities, the IoT system is unable to store a huge amount of sensor-generated data and processing of this data is also not up to the mark. Therefore, the IoT system must need a helping hand that can overcome these limitations. Cloud computing has a pool of resources, that consists of infinite storage capacity, immense computing power and network bandwidth, in addition to many others, that can help IoT system to overcome the above-discussed issues. Cloud computing resources are elastic in nature, they can expand and reduce according to the IoT environment needs. Cloud-based big data analytics can also help in analysing sensor data [39]. Integration of cloud computing with IoT system helps IoT-based applications to become more efficient and reliable. Integration of IoT with cloud computing has many benefits. These benefits are given below.

*5.4.1* ***Data Transmission***. Cloud-based IoT paradigm provides data exchange facilities. IoT applications can transfer data from one node to another, communication between the nodes and data distribution among data nodes, all at very low cost. Cloud computing provides economical and efficient solution for data connection, control and communication using integrated applications.

*5.4.2* ***Storage***. IoT environment consists of many connected devices and sensors, which generate huge amount of data in real time. The IoT local storage is not capable enough to store such amount of data. IoT devices also generate unstructured and semi-structured data. The traditional databases are not capable to store data of such high variety. Cloud computing solves this data storage problem for IoT systems. Cloud Computing has a collection of commodity computers, which have huge amount of integrated storage. IoT devices can store their data onto the Cloud and can access the same from anywhere via Internet. This huge storage of data can be furthermore used for analytics and system improvements.

*5.4.3* ***Processing***. The processing power of IoT devices are very limited. Therefore, they cannot process huge volumes of data produced by millions of connected smart devices. Cloud computing provides computing power to the IoT systems by making use of virtualization technique that divides the physical machine into number of virtual machines. IoT devices can hire these virtual machines for their application execution on pay per use basis. Cloud computing integrates with IoT system enables the application users a low-cost processing power with high revenue.

*5.4.4* ***Modern Capacities***. IoT system has various heterogeneous connected devices. These devices use different protocols and mechanisms. Therefore, the coordination between these heterogeneous devices are very difficult. Moreover, achieving optimum reliability and efficiency can also be challenging. Cloud computing can solve these issues of heterogeneity of devices, by its elastic and easy to use nature. Integration of cloud computing with IoT gives reliability, scalability, security and efficiency to user's applications.

*5.4.5* **Cloud of Things Models**. Cloud computing has three main deployment models namely, SaaS, IaaS and PaaS. Integration of cloud with IoT gives birth to various new deployment models, which are listed below:

- Sensing as a Service (SaaS) provide services to access IoT sensors data.
- Ethernet as a Service ( EaaS ) provides internet connection services to the IoT devices.
- Sensing and Actuation as a Service ( SAaaS ) provides control services over the sensors and devices automatically.
- Identity and Policy Management as a Service ( IPMaaS ) provides identification of connected devices and access policies.
- Database as a Service ( DBaaS ) provides database services for data storage and management.

## 5.5   Cloud-based IoT Architecture :

Basically, there are three levels of IoT architecture: Applications Layer, Network layer and Perception layer. Perception Layer collects the data from the physical environment and sends to the Network layer. Network Layer receive the data from Network Layer and send to the Applications where it applies for various interesting applications [73] [76]. The cloud-based IoT architecture is shown in Fig. 13.

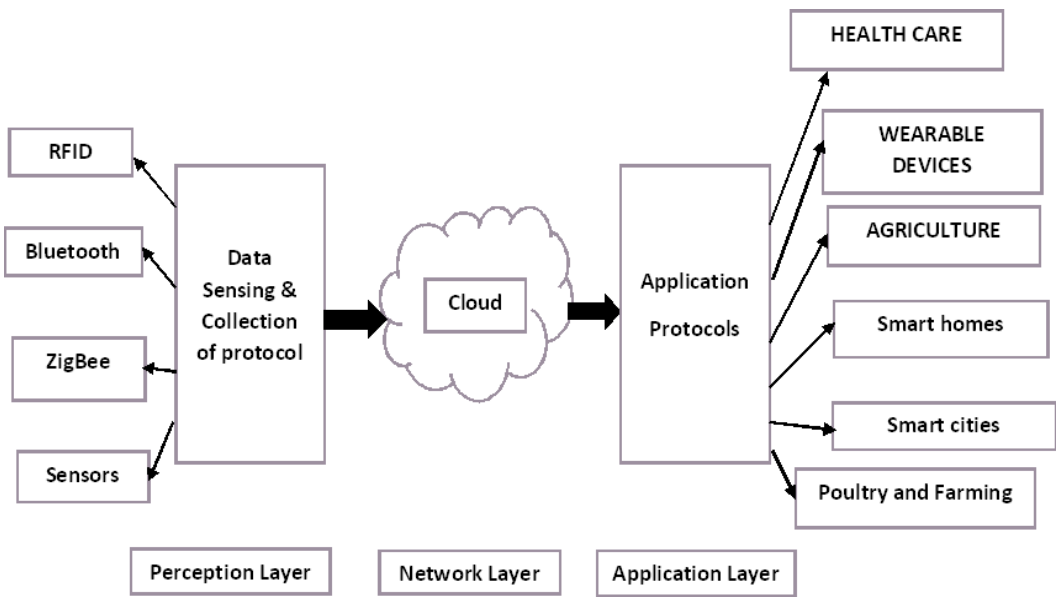

Fig. 13. **Cloud of Things Architecture**

## 5.6   Cloud-based IoT Application

IoT and cloud are inextricably linked. The data collected by the sensors is quite large in case of an industrial IoT application and a gateway is unable to process and store it. This data should be stored in a secure database and processed in a convenient and scalable manner. This is where IoT and cloud come into picture. The cloud-based IoT approach presented various applications and keen administrations, influencing end clients consistently. Various Cloud-based IoT applications have highlighted in Table 5

Table 5. **Various Cloud-based IoT Applications**

| Application | Description |
| --- | --- |
| Healthcare [78] | The healthcare sector can be more advanced due to Cloud-based IoT architecture. Sensor networks are used by the hospitals to gather patient health-related data using special sensors. |
| Smart cities [33] | IoT enabled Smart city projects to make human life easier and more efficient. Many subsystems contribute to the development of the smart city. energy management, transportation management, smart dustbin and smart surveillance are the popular inventions contribute to the development of smart cities. |
| Smart Home [67] | Smart Home has the basis of smart mobile phone model as a smartphone can do multiple tasks same as in smart homes the things are connected using IoT network system and do tasks automatically in a customized manner according to the user's choice. |
| Smart Agriculture [41] | Now a day's agriculture is a growing application of IoT system. The farmers can use their mobile phones for the surveillance of their field from anywhere and take appropriate action regarding watering, insect detection and pesticide the field to protect from harmful insects' diseases. |
| Industrial Internet of Things(IIoT) [55] | Industrial Internet of Things (IIoT) is a new smart industry management method using smart devices, sensors and computing system. IIoT empowering the industries to get data about the inventory and manufacturing unit in run time and sensor-enabled alarming system alert the employees. The data captured by the sensors stored and analysed for future plans and directions. |
| Connected Cars [56] | A sensor-enabled car that optimizes the operations of driving and gives comfort to the driver, navigate to the destination, protect from traffic roadblocks and accidents. Connected cars are very helpful in the development of smart city framework. Connected cars can improve on-road traffic and reduce transportation cost efficiently. |
| Wearables Devices [77] | Today health and fitness are the main concern of human, the wearable devices enabled with sensors sense the human daily work-activities, heart pulse and many more data to help people to get information about their health. These devices are easily connected with mobile phones via Bluetooth. The data stored in mobile locally as well as on Cloud storage for analysis and finding the changing pattern of data. Most of the hospitals used these wearable devices for patients to monitor their pulse rate, blood pressure and all. |
| Poultry and Farming [57] | Besides the wearable devices for human health monitoring, the IoT is also used to monitor the health of cattle. IoT devices and sensors are also widely used in poultry and farming. |

| *Table 5 continued* | |
| --- | --- |
| **Energy Engagement [30]** | The energy consumption by the industries as well as the residential area is grown day by day. The huge amount of energy usage and its solution for efficient energy consumption are the major objectives of the government. The IoT sensors enabled smart grids are used to collect data about the consumption automatically and analyse the cause of huge energy consumption and how to improve it efficiently to ensure economic benefits to the consumers and suppliers. |
| **Smart Retail [59]** | Smart retail is one of the most attracting applications of IoT, which enable an automated selling environment for the customer as well as the retailer. The smart devices and sensors are connected to the selling products and well as the retail smart cart. User can add items into the cart and billing amount is calculated automatically and paid by the debit or credit cards of the customer automatically. |

## 6  THE FUTURE OF IOT AND ITS INTERACTION WITH BLOCKCHAIN

This section focuses on the integration of IoT with Blockchain and its various benefits in a systematic manner. The complementary approaches and issues regarding Blockchain and IoT integration are also identified in this section. Moreover, this section addresses, how Blockchain face s challenges of IoT convergence. IoT and Blockchain are going to be technologies that change the world, right at the beginning of their adoption curves.

Furthermore, how Blockchain has provided security and privacy features to enable IoT applications more reliable and secure. The integration of IoT with Blockchain gives more reliability in cloud-based IoT architecture applications.

IoT is an Internet interconnection of computer devices integrated into day to day objects, allowing them to send and receive data. It is a concept used in this work and in personal lives to describe the ongoing proliferation of online data collection tools whereas Blockchain is an encrypted, decentralized database file system designed to create tamper-proof records in real-time [27]. Block chain technology is used for extracting the information and storing this information from various IoT devices in the form of blocks to ensure transparency among the globally located users [69]. Various security techniques and approach es have been highlighted in the last decade. Blockchain plays a vital role in securing many IoT-based applications [66]. The decentralized architecture of Blockchain-based IoT system can offer security and many other benefits. The benefits of the integration of IoT with Blockchain have been discussed in the next section.

### 6.1  Benefits of Integrating IoT with Blockchain:

IoT integration with Blockchain provides many advantages to both revolutionize technologies. Blockchain incorporation with IoT has many advantages as shown in Fig.14.

The integration of IoT and Blockchain [76] [38] [40] has following benefits:

#### 6.1.1 Data Integrity

Blockchain technology is a peer-to-peer based network in which the same record of data has been copied to every node in the network. All the transactions in Blockchain are validated by the private key. Every node sends its private key to another node for validation. A nonce [10] value has been found by all the minor nodes in the network for validation. The node finds the nonce value first will announce as the winner node and will have the right to validate and get rewarded.

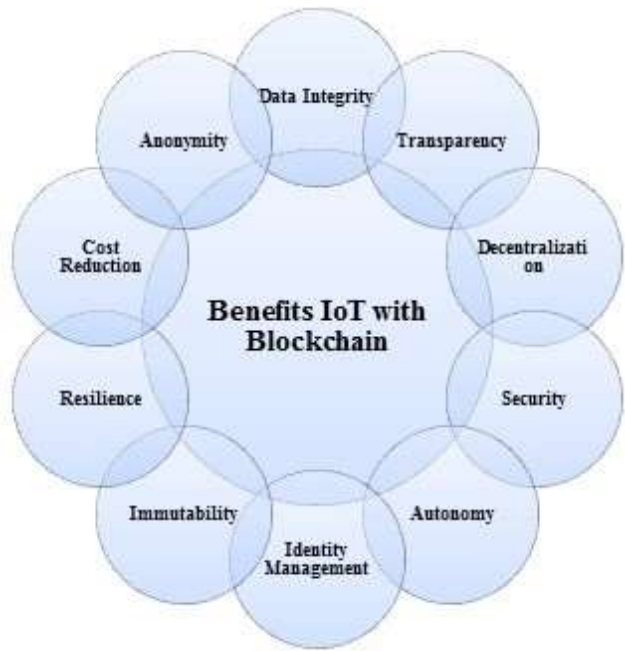

Fig. 14. **Benefits of Integration of IoT and Blockchain**

The newly added block can not be modified or deleted in the Blockchain and all other nodes in the network will be informed for this newly created block [8].

### 6.1.2 Transparency

In a Blockchain, each participant has its own ledger and can check all the transactions. All the transaction data has been protected by the private key of each node. Therefore, all the participants can see the transaction, but they cannot modify other ' s transaction data due to private key protection mechanism. In a IoT network system, all devices share their data with each other. Blockchain-integrated IoT system can provide secure data communication among different network devices [47].

### 6.1.3 Security

Security is the main concern in IoT that can be enhanced significantly with Blockchain integration. Subsequently, Blockchain offers security against the malicious actions by using public key infrastructure. Transactions are shared by all nodes in the IoT network and protected by the private key authentication [48].

### 6.1.4 Decentralization

Blockchain empowers decentralized and distributed features in which central authority is not needed to manage and execute operations. This feature offers ' to and fro' communication between different nodes of the network. To validate a transaction or set the rules for approving a transaction, no overarching authority exists. Thus, a massive amount of legitimacy is included as most of the network users must reach an agreement to verify transactions [71].

### 6.1.5 Autonomy

Artificial intelligence-enabled autonomous hardware and resources as a service are provided by the Blockchain technology. Smart car and automobiles need Blockchain technique for secure and mutable data security to prevent cyber-attack. Since security is an important aspect of autonomous devices, Blockchain with artificial intelligence and machine learning techniques enable the development of new autonomous IoT applications.

### 6.1.6 Identity Management

The IoT network is witnessing identity theft with increasing number of devices and artefacts, allowing malicious entities to collect and exploit sensitive information from legal entities. Blockchain convergence with IoT can solve this problem, as this technology can easily manage the node credentials and identity of the user [79]. Furthermore, Blockchain can provide IoT applications with secure decentralized encryption and system authorization.

### 6.1.7 Immutability

Blockchain offers a disruptive feature in which if a participating user wants to add a transaction, the transaction is only added to the block. If it is verified by most of the nodes involved in the Blockchain network, which is performed using consensus algorithm, the automated validation is conducted efficiently so that each client produces a fast and secure ledger for transactions and blocks which is substantially tamper-proof. The conventional security solutions applied to the centralized network in the IoT system do not protect data integrity, but with the immutability of the Blockchain, data integrity is ensured [63].

### 6.1.8 Resilience

The elastic feature of Blockchain provides a shield to process trans actions in distributed locations. Each client in participation holds the original copy of the transaction ledger, which is mandatory in Blockchain. Thus, it can handle the attacks because the reliability is high on Blockchain. If one node fails, data can be accessed from the other nodes in the network. Cross platform data exchange in IoT applications is easy and effective due to copy of data but processing of data is difficult.

### 6.1.9 Anonymity

Both the consumer and the vendor use anonymous and SHA 160 enabled public key as address numbers to process the payment, which preserves their private identity. The anonymous participation of the user has some drawback, because anonymous participation may lead to illegal online market whereas it can be beneficial for an anonymous voting system.

### 6.1.10      Cost Reduction

Due to the massive requirement of infrastructure and maintenance cost of the centralized and cloud-based technologies, the IoT systems are very expensive. The communications among zillions of IoT systems are up and running will significantly increase these costs. The Blockchain-enabled IoT system can slightly reduce the cost of communications. The Benefits of Integration of IoT and Blockchain is given in fig 14.

## 6.2   Blockchain as solutions for different IoT challenges

IoT systems are mainly suffering from security aspects. Blockchain plays an important role to mitigate the challenges associated with security. Blockchain can manage and control security aspects of IoT systems effortlessly by smart contacts. Table 6 explores and outlines most of the inherent capabilities of Blockchain that can be of enormous benefit to IoT in general, and I

## 6.3   Cloud of things (CoT) — Challenges and Open Research Issues

The integration of IoT with Cloud Computing (Cloud of Things) may be hindered by many challenges and open issues. The identified challenges and issues are:

• **Security and Privacy** : In the implementation of Internet of things (IoT) — based cloud infrastructures, security and privacy are the main concern s. Cloud-based IoT allows real-world information transmission to the cloud. In addition, an important issue still not resolved is how effective authorization rules and policies should be developed even though confirming that only authorized users have access to data. It is essential when it emanates to protecting the confidentiality of users and preserving data integrity. In fact, as sensitive IoT systems migrate to the cloud, some issues arise due to absence of transparency of service level agreement, provider details and the physical location of the software. Multi-tenancy can also cause sensitive data leakage. Public-key cryptography cannot necessarily be applied to all levels due to various IoT's processing power restrictions. In contrast, major flaws are also troublesome, including user hijacking and virtual machine rescue. The attack and suggestion is given by the researchers [31], they try to discover the main cause for attacks and suggested a solution which is based on Network Neural and Cloud TraceBack (CTB).

Table 6. **Blockchain Solutions for IoT challenges**

| IoT challenge | Blockchain as solutions for different IoT challenges. |
|---|---|
| **Security** | The main concern of Blockchain is to provide security. It offers trust and unchangeable system architecture for IoT system. It also provides a private key authentication mechanism for modification in the block by any network node [53]. The key features of Blockchain are data integrity, privacy and confidentiality. |
| **System single-piont failure** | Blockchain is a decentralized and distributed technology for data security that can prevent the single point failure of centralized IoT system due to its distributed property [46]. |
| **Authority of the third party** | Blockchain is fundamentally immutable database ledger, distributed and decentralized framework for different IoT devices. Therefore, no third-party authority is required between the communications nodes. |
| **Address space** | A huge amount of heterogeneous IoT smart devices are connected in the IoT network infrastructure. The address-space of Blockchain is 160-bit which can configure approximately 1,446,448 offline IoT devices. Therefore, Blockchain can connect near about 4.3 billion more IoT devices than IPv4 and IPv6. |
| **Malicious Vulnerability** | Due to the transparent and unchallengeable feature of Blockchain, the malicious attack in the IoT network can easily be detected and prevented. Update of any information in a block can be done only if all participating nodes in the IoT network authenticate with their private key authentication. |
| **Identity and possession** | The Blockchain-enabled network can provide secure, approved identify registration, monitoring, and tracking of possession. It can also be used to authorize IoT devices and assign them identities [60]. |
| **Access control and Authentication** | The contracts for smart Blockchain can provide decentralized encryption and logic that enables effective IoT authentication. |
| **Elasticity** | IoT organizations can accomplish multiple goals through various business and open-source options for Blockchain without investing massively in research and development. |
| **Data Integrity** | An immutable and unchangeable transaction ledger of Blockchain delivers secure communication data sharing and efficient monitoring of data among all IoT devices. Data integrity has been achieved by the Blockchain network in IoT applications cryptographically. |
| **Capabilities and Economic issues** | Although the Blockchain does not require a centralized server, the communication among the IoT devices are secure. The devices can exchange data and take action automatically through smart contracts. |

• **Ipv6 addressing strategies** : Internet is one of the key components of IoT, which has been influenced by IPv4. Constrained Applications Protocol (CoAP) technology can communicate directly with devices integrated by the Internet. The subsequent development of these technologies includes the removal of processes for Network Address Translation (NAT) to target the growing IoT system or network with a unique IP address. IPv6 is designed to address restrictions related to IPv4 by using a 128-bit IP address and has many benefits, such as the wide range of Internet-connected devices, end-to-end access and the REST interface agreement. On integrated IoT devices, IPv6 implementations can be performed using IP specifications 6LoWPAN and Zig Bee that ha s yet to be extended on the platforms of many areas. The IoT network is using IPv6 from human-initiated networks.

• **Interoperability** : Interoperability is a major issue in c loud-based IoT systems due to heterogeneous and autonomous nature of IoT devices. Many types of research have been done in this area to provide the solution to this issue in the last few years. The proposed methods provide different solutions that arise the problem of interoperability due to heterogeneity in c loud platform and applications. The interoperability may also lead the inability in cross-platform and cross-domain applications development [68].

• **Intelligent Analytics** : The centralization of real-time data from heterogeneous entities into the cloud allows improved decision-making capability using advanced intelligence processing and integration processes. An n umber of challenges that prevent intelligent analytics such as erratic machine activity during an accident, Classic analytics systems, poor implementation of the latest technology due to high cost, lack of skilled data mining experts, algorithms, artificial intelligence, and sophisticated event management. Recent efforts have been made in this area [28], Enhancing intelligence is an open challenge in this context.

• **Integration methodology** : The requirement for interoperability cannot be ignored by integrating current and future smart cyber-physical systems into a fully realized Internet of Things (IoT). The lack of IoT standards and their inherent demand for complex methodologies to support the development and integration of interoperable and heterogeneous IoT systems, there are no methodologies for integrating IoT systems [72].

• **Heterogeneity** : One of the main challenges of IoT and cloud computing implementation is the wide diversity of existing and potential applications for new and modified software, operating systems, platforms or services. Therefore, as end-users implement multi-Cloud solutions, the heterogeneity problem can be compounded, and applications rely on multiple providers to enhance scalability and efficiency.

• **Standardizations** : A growing number of researchers identify the lack of standards as a major issue for CloudIoT model. Although the technological community has proposed several recommended standardization options for the application of IoT and cloud technology, standard protocols, interfaces and APIs are needed to enable heterogeneous smart things to be interconnected and novel services to be built which make up the IoT-based Cloud paradigm.

- **Fog Computing** : Fog computing is a model that applies cloud-based services to the network's edge. Fog delivers application services to clients related to the cloud. In general, Fog can be considered as a cloud system extension, serving as a bridge between the cloud and edge of the network. It works for application resistant to latency which requires additional nodes to fulfill their latency requirements. Though processing, networking and computation are the key assets of Cloud and Fog. The fog has certain functions that provide geo-distribution and low latency, such as edge detection and location recognition. There are also large clusters, in addition to Cloud which facilitates real-time connectivity and flexibility.
- **Cloud Capabilities** : Security is regarded as a major issue in the cloud-based IoT model for any networked system. There are more possibilities for attacks on both the IoT and the cloud side. Data privacy, anonymity and authentication can be ensured by encryption in the IoT context. In contrast, internal threats could not be conquered, and IoT on devices with limited capabilities is also difficult to use.
- **SLA Implementation** : IoT-based cloud applications allow the transmission and storage of data generated in compliance with application-dependent constraints, which may be problematic in certain situations. It might not always be possible for a single vendor to ensure a certain amount of QoS for cloud services, therefore it may be necessary to depend on various providers of cloud SLA violations.
- **Big data** : An expected 50 billion units are to be networked by 2020, and the massive amount of data generated by them needs to be delivered, collected, retrieved and analysed. The ubiquity of mobile devices and the pervasiveness of sensors practically requires modular systems. Therefore, various Cloud service providers might be needed to prevent SLA file breaches. Nonetheless, the flexible option of the most suitable blend of cloud services remains an open issue due to the time, expense and QoS complexity maintenance support.
- **Power and energy efficiency** : Today's IoT-based applications provide frequent data transferred from IoT entities to the cloud that quickly absorb node power. Therefore, energy efficiency is still a major concern in the area of data processing and transmission.
- **Performance** : Many technologies of cloud computing and IoT implementation have specific performance specifications or QoS criteria at different scales (e.g. connectivity, processing, and storage) and standards may not be easily achieved in certain situations.
- **Reliability** : In mission-critical applications, cloud computing and IoT convergence generally pose reliability issues, e.g. in the area of smart mobility, automobiles are often moving, and automobile networking and connectivity are often sporadic and ineffective. A variety of problems relevant to system malfunction or not always reachable systems are raised in a resource-controlled environment.
- **Monitoring** : Monitoring is a crucial activity for batch processing, resource management, SLAs, reliability and security, and troubleshooting in cloud environments. Therefore, the Cloud-based IoT strategy has the same cloud monitoring criteria, even though some specific problems remain which are caused by the speed, scale, and complexity of the IoT.

## 7   CONCLUSION

Cloud computing and IoT integration is indicative of the subsequent big leap in the world of Internet. The new class of applications that come into existence with this combination are known as the IoT Cloud. This research paper surveys the different aspects of Cloud, IoT and the benefits and challenges that exist in creation of a synergistic approach. Besides this, it also elaborates on the integration of Blockchain for mitigating the security-related challenges associated with using Cloud and IoT together for varied applications. Cloud-IoT is opening new avenues for business and research. We hope that this combination will reveal a new paradigm for the future of multiple networks and an open service platform for users.

## ACKNOWLEDGMENTS

This work was supported by Big Data, Cloud Computing and IoT Laboratory, Department of Computer Science, JMI, a grant from "Young Faculty Research Fellowship" under Visvesvaraya PhD Scheme for Electronics and IT, Department of Electronics & Information Technology (DeitY), Ministry of Communications & IT, Government of India.